\documentclass[a4paper,11pt]{article}
\usepackage[no-natbib-sort]{pos}
\usepackage{tikz}
\usepackage{nicematrix}
\usepackage{caption}
\usepackage{subcaption}
\usepackage{multirow}
\usepackage{cleveref}

\definecolor{rwthblue}{HTML}{00549F}
\definecolor{rwthblue50}{HTML}{8EBAE5}
\definecolor{rwthblue25}{HTML}{C7DDF2}
\definecolor{rwthblue10}{HTML}{E8F1FA}
\definecolor{rwthgreen}{HTML}{57AB27}
\definecolor{rwthred}{HTML}{CC071E}

\def\be{\begin{equation}}
\def\ee{\end{equation}}
\def\bea{\begin{equation}\begin{aligned}}
\def\eea{\end{aligned}\end{equation}}

\newcommand{\history}{\texttt{history}}
\newcommand{\feyngame}{\texttt{FeynGame}}

\newcommand{\vhnnlo}{\texttt{vh@nnlo}}
\newcommand{\mcfm}{\texttt{MCFM}}

\newcommand{\pp}{pp}
\newcommand{\qqp}{qq^{\prime}}
\newcommand{\qqb}{q\overline{q}}
\newcommand{\qg}{qg}
\newcommand{\ggc}{gg}
\newcommand{\vh}{VH}
\newcommand{\wh}{WH}
\newcommand{\zh}{ZH}

\Crefname{figure}{Figure}{Figures}

\title{\texttt{history} in the making: A tool for NNLO cross sections}

\author*[a,b]{Lukas Simon}
\author[b]{Sven Yannick Klein}

\affiliation[a]{Laboratoire de Physique Th\'eorique et Hautes Energies (LPTHE), UMR 7589, Sorbonne Universit\'e et CNRS, 4 place Jussieu, 75252 Paris Cedex 05, France}

\affiliation[b]{Institute for Theoretical Particle Physics and Cosmology, RWTH Aachen University, Sommerfeldstraße 16, 52074 Aachen, Germany}

\emailAdd{lsimon@lpthe.jussieu.fr}
\emailAdd{yklein@physik.rwth-aachen.de}

\abstract{In these proceedings, we report on our progress in developing the \history{} framework, which aims to implement the fully-local Nested Soft-Collinear infrared subtraction scheme for the automated phase-space integration of color-singlet production processes in hadronic collisions at NNLO accuracy. We validate our implementation for quark-antiquark-initiated processes and demonstrate a first application of the tool by predicting a novel observable for the inclusive process $\pp \to \zh+X$, which may offer sensitivity to potential effects of new physics.}

\FullConference{17th International Symposium on Radiative Corrections: Applications of Quantum Field Theory to Phenomenology (RADCOR2025)\\
5-10 October 2025\\
Puri, India\\}

\begin{document}
\maketitle

\section{Introduction}

Originally designed as a discovery machine, the Large Hadron Collider (LHC) has also proven to be an ideal facility for precision measurements. In the absence of clear signals of physics Beyond-the-Standard Model (BSM), high-precision tests of the Standard Model (SM) have become a central objective of particle physics, with the hope that small deviations from SM predictions may offer hints of new physics. Achieving this goal requires not only advanced experimental techniques but also highly accurate theoretical predictions of the relevant observables, including higher-order perturbative corrections.

While leading-order (LO) and next-to-leading-order (NLO) corrections in the perturbative expansions of the strong and electroweak couplings, $\alpha_s$  and $\alpha$, are fully automated, next-to-next-to-leading-order (NNLO) computations remain challenging and are performed on a process-by-process basis. Nonetheless, NNLO accuracy is essential to match the precision reached in modern experimental analyses. Owing to significant theoretical and computational advances, the automation of NNLO calculations is becoming increasingly realistic. This progress is driven both by a deeper understanding of the structure of multi-loop Feynman integrals and improved methods for evaluating scattering amplitudes, as well as by the development of efficient infrared (IR) subtraction schemes that enable automated phase-space integration for complex processes. A variety of IR subtraction schemes at NNLO have been proposed over the past two decades~\cite{Catani_2007,Gaunt15,Ridder_2005,Czakon_2010,Cacciari_2015,DelDuca:2016ily,Caola_2017,Herzog_2018,Magnea_2018,Anastasiou:2022eym}.

\begin{figure}
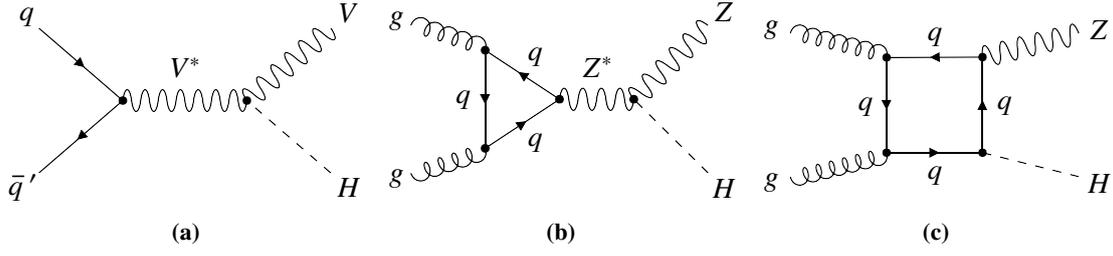

\center
\begin{subfigure}[b]{0.32\textwidth}
\def\svgwidth{\textwidth}\input{./figures/qqVH.tex}
\caption{}\label{fig:dy}
\end{subfigure}
\begin{subfigure}[b]{0.32\textwidth}
\def\svgwidth{\textwidth}\input{./figures/ggZH_tri.tex}
\caption{}\label{fig:tri}
\end{subfigure}
\begin{subfigure}[b]{0.32\textwidth}
\def\svgwidth{\textwidth}\input{./figures/ggZH_box.tex}
\caption{}\label{fig:box}
\end{subfigure}
\caption{Feynman diagrams for associated Higgs production. Panel (a) shows the Drell-Yan mechanism, initiated by quark-antiquark annihilation with an intermediate vector boson $V$, where $V$ denotes either a $W$ or a $Z$ boson. Panels (b) and (c) depict the loop-induced gluon-fusion contributions to $\zh$ production. The diagrams are generated with \feyngame~\cite{feyngame,feyngame2,feyngame3}.}
\end{figure}

To predict fully-differential cross sections at NNLO, we are developing the tool \history{}~\cite{Simon:2023zuk}. It is based on the fully-local Nested Soft-Collinear (NSC) subtraction scheme~\cite{Caola_2017,Devoto_2023,Devoto:2025kin,Devoto:2025jql} that builds upon the \texttt{STRIPPER} approach~\cite{Czakon_2010,Czakon_2011,Czakon_2015}. Our framework is designed to implement the NSC algorithm for arbitrary color-singlet processes at NNLO QCD accuracy~\cite{Caola_2019}. In its current form, \history{} supports processes that are initiated by quark-antiquark annihilation at Born level, while the extension to gluon-fusion processes is underway. Although the framework is in principle applicable to any color-singlet process with a quark-antiquark initial state, the corresponding matrix elements must still be interfaced on a process-by-process basis. As a result, the NNLO processes currently available are limited to the associated production of a Higgs boson and a vector boson produced via the Drell-Yan mechanism. The corresponding Born-level Feynman diagram for this process is shown in \cref{fig:dy}, where a quark and an antiquark annihilate into a virtual vector boson $V^\ast$, which then radiates a Higgs boson, $\pp \to V^\ast+X \to \vh+X$. Nevertheless, extensions to additional processes are straightforward to incorporate once the relevant two-loop amplitudes are available.

Associated Higgs production is itself an interesting process to investigate, particularly when comparing the Drell-Yan production mode with the loop-induced gluon-fusion contribution to $\zh$ production. Even though the gluon-induced channel is formally suppressed by a factor $\alpha_s^2$, it contributes roughly $20\%$ of the total $\pp \to \zh+X$ cross section in the SM. At $\mathcal{O}(\alpha_s^2)$, the process is described by a triangle and a box diagram, shown in \cref{fig:tri,fig:box}, respectively. The loop-induced channel plays an important phenomenological role due to its sensitivity to potential new-physics effects. The triangle diagram may receive contributions from BSM particles circulating in the loop or appearing in the intermediate propagator, while the box diagram can be affected both by new particles in the loop and by modified Yukawa interactions, as predicted in various BSM scenarios. Extracting the gluon-induced component from measurements of the $\zh$ cross section would therefore be highly desirable, as it could provide evidence for such new-physics effects.

The associated production of a Higgs boson and a vector boson has been extensively studied in the literature. The total inclusive cross section at NNLO accuracy was computed two decades ago~\cite{Brein:2003wg}, and fully-differential predictions at the same precision, based on phase-space slicing methods, have been available for more than ten years~\cite{Ferrera:2011bk,Ferrera:2013yga,Ferrera:2014lca,Campbell:2016jau}. The loop-induced process $\ggc \to \zh+X$ was computed at $\mathcal{O}(\alpha_s^2)$ in the late 1980s and early 1990s~\cite{Barger:1986jt,Dicus:1988yh,Kniehl:1990iva}, the $\mathcal{O}(\alpha_s^3)$ corrections including exact top-quark mass effects were published roughly five years ago~\cite{Chen:2020gae}.

\section{Validation of the implementation}

\begin{figure}
    \centering
    \begin{subfigure}[c]{0.3\textwidth}
    \begin{small}
         \begin{NiceTabular}{ l | r }[color-inside]
 \arrayrulecolor{rwthblue}
    \rowcolor{rwthblue}
	 \multirow{2}{*}{\phantom{......}\textcolor{white}{channel}\phantom{......}} & \phantom{..}\textcolor{white}{\history{} $\mathrm{[fb]}$}\phantom{..} \\ \rowcolor{rwthblue}
	 & \phantom{..}\textcolor{white}{\vhnnlo{} $\mathrm{[fb]}$}\phantom{..} \\ \rowcolor{rwthblue50}
        \multirow{2}{*}{$\pp \to \zh+X$} & $848.274(35)$ \\ \rowcolor{rwthblue50} & $848.269(76)$ \\ \hline\rowcolor{rwthblue10}
        \multirow{2}{*}{$\qqp \to \zh+X$} & $966.131(12)$ \\ \rowcolor{rwthblue10} & $966.102(71)$ \\ \hline\rowcolor{rwthblue10}
        \multirow{2}{*}{$\qg \to \zh+X$} & $-120.332(10)$ \\ \rowcolor{rwthblue10} & $-120.346(26)$ \\ \hline\rowcolor{rwthblue10}
        \multirow{2}{*}{$\ggc \to \zh+\qqb$} & $2.51256(15)$ \\ \rowcolor{rwthblue10} & $2.51319(50)$
 \end{NiceTabular} 
    \end{small}
    \end{subfigure}\hfill
    \begin{subfigure}[c]{0.64\textwidth}
    \input{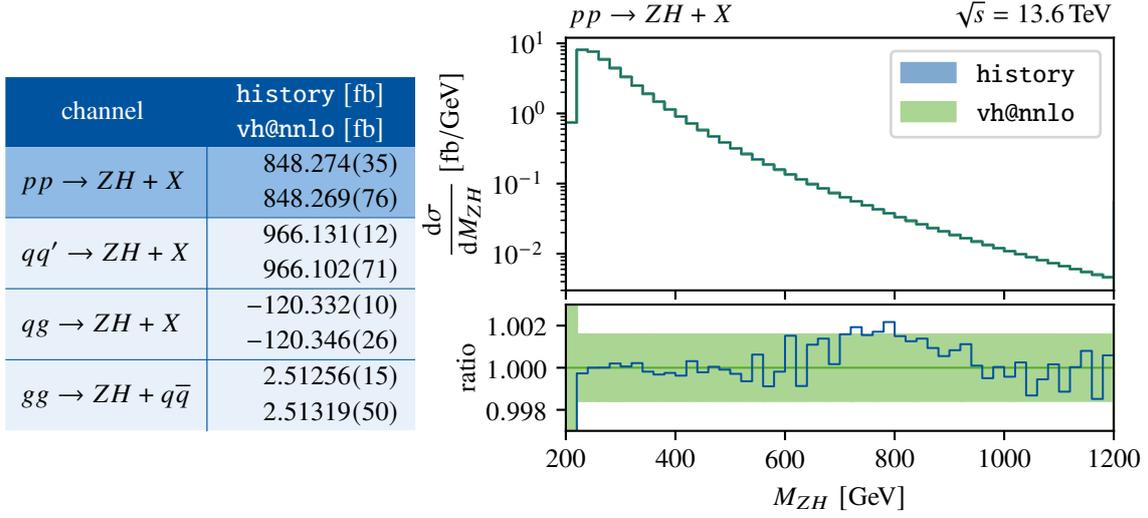}
    \end{subfigure}
    \caption{\textit{left}: Total inclusive NNLO cross section for $\pp \to \zh+X$ via the Drell-Yan mechanism and the contributions from the individual partonic channels compared between \history{} (upper values) and \vhnnlo{} (lower values). Statistical Monte Carlo uncertainties on the last two digits are quoted in parentheses. \textit{right}: Invariant mass spectrum for $\pp \to \zh+X$ at NNLO precision. The upper panel shows the spectra computed with \history{} (blue) and \vhnnlo{} (green). The lower panel displays their ratios normalized to the \vhnnlo{} distribution. The shaded bands indicate the statistical Monte Carlo uncertainties.}
	\label{fig:MZH}
\end{figure}

Before turning our attention to the phenomenology of associated Higgs production, we first validate our implementation for color-singlet processes that are initiated by quark-antiquark fusion at Born level. To this end, we benchmark \history{} against \vhnnlo{}~\cite{Brein:2012ne,Harlander:2018yio}. Using both frameworks, we compute the total inclusive cross section for associated Higgs-$Z$ boson production in proton-proton collisions at $\sqrt{s}=13\,\mathrm{TeV}$ in the Drell-Yan production channel at NNLO QCD accuracy. The numerical results, summarized in the table on the left of \cref{fig:MZH}, reveal perfect agreement between \history{} and \vhnnlo{} within the quoted Monte Carlo (MC) uncertainties for all individual partonic channels as well as for the total cross section. This excellent level of agreement provides a first and crucial validation of our implementation, demonstrating that the local NSC subtraction counterterms correctly cancel the IR singularities of the real-emission contributions up to NNLO, while the integrated counterterms precisely reproduce the oversubtracted terms.

A second benchmark is presented on the right of \cref{fig:MZH}, where we compare the invariant-mass distributions of the final-state Higgs-vector boson system as obtained with the two codes. This comparison probes the stability of the subtraction method across a huge range of the phase space and therefore represents a more stringent test than the total inclusive cross section alone. Also in this case, we observe agreement at the permille level, with all individual bins consistently matching within the respective MC uncertainty bands. The consistency of the two predictions across the full kinematic range further confirms the robustness and correctness of our implementation of the NSC subtraction formalism in \history{}.

\section{Associated Higgs production and new physics}

Having the \history{} framework at hand, we are able to calculate fully-differential cross section for the production of a Higgs boson together with either a $W$ or a $Z$ boson. This enables the study of the $R^{\vh}$ double ratio that has been proposed in ref.~\cite{Harlander:2018yns}. This observable allows the separation of the gluon-induced contribution, $\ggc \to \zh+X$, from the Drell-Yan component, $\qqb \to Z^\ast+X \to \zh+X$. While the former mechanism is potentially very sensitive to a wide variety of BSM scenarios and thus represents an ideal channel to probe experimentally, the latter is the dominant production mode and constitutes an irreducible background in experimental analyses.

The $R^{\vh}$ double ratio is defined as
\be\label{eq:double_ratio}
R^{\zh}(x)=\dfrac{R^{\vh}(x)}{R^{\vh}_\mathrm{DY}(x)}\,,
\ee
where the numerator is the ratio of the experimentally measurable (differential) cross sections for the processes $\pp \to \zh+X$ and $\pp \to \wh+X$,
\be\label{eq:ratio_exp}
R^{\vh}(x) = \dfrac{\mathrm{d}\sigma_{\pp\to\zh+X}/\mathrm{d}x}{\mathrm{d}\sigma_{\pp\to\wh+X}/\mathrm{d}x}\,,
\ee
while the denominator contains the theoretically predicted (differential) cross sections of the same processes, but only including the contributions from the Drell-Yan production mechanism,
\be\label{eq:ratio_theo}
R^{\vh}_\mathrm{DY}(x) = \dfrac{\mathrm{d}\sigma_{\pp\to\zh+X}^\mathrm{DY}/\mathrm{d}x}{\mathrm{d}\sigma_{\pp\to\wh+X}^\mathrm{DY}/\mathrm{d}x}\,.
\ee
Here, the symbol $x$ denotes an IR-safe observable. The use of ratios in both the numerator and denominator of \cref{eq:double_ratio} yields a particularly precise observable. Experimentally, the ratio in \cref{eq:ratio_exp} benefits from partial cancellations of systematic uncertainties common to both the $\zh$ and $\wh$ measurements, such as the luminosity uncertainty. Similarly, the theoretical uncertainties from scale variations in \cref{eq:ratio_theo} are strongly correlated between the two processes, leading to a highly precise prediction.

\begin{figure}
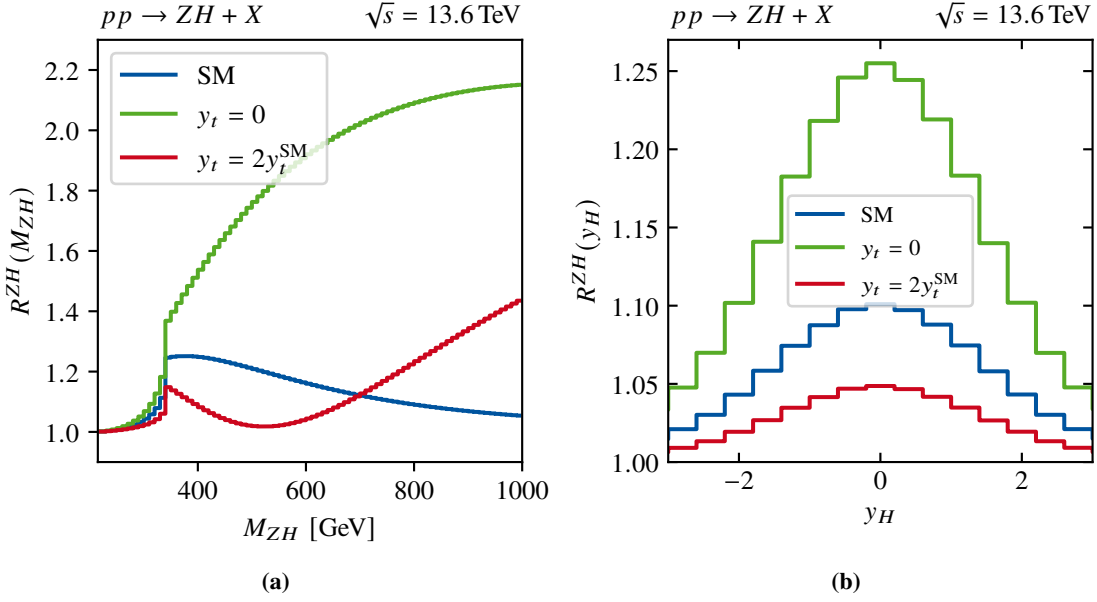

    \centering
    \begin{subfigure}[b]{0.49\textwidth}
    \input{figures/Fig2a.pgf}
    \caption{}
    \label{fig:RVH_a}
    \end{subfigure}
    \begin{subfigure}[b]{0.49\textwidth}
    \input{figures/Fig2b.pgf}
    \caption{}
    \label{fig:RVH_b}
    \end{subfigure}
    \caption{The $R^{\vh}$ double ratio as function of (a) the invariant mass $M_{\zh}$ obtained with \vhnnlo{}, a reproduced plot from ref.~\cite{Harlander:2018yns}, and (b) the rapidity of the Higgs boson $y_H$ obtained with \history{}. The double ratios are computed in the SM (blue), the SM without a top-to-Higgs Yukawa coupling (green), and the SM with a doubled top-to-Higgs Yukawa coupling (red).}
	\label{fig:RVH}
\end{figure}

Based on computations with \vhnnlo{} interfaced with \mcfm{}~\cite{Campbell:2016jau,Boughezal:2016wmq}, ref.~\cite{Harlander:2018yns} demonstrated that the invariant mass of the $\vh$ system, $M_{\vh}$, and the transverse momentum of the Higgs, $p_{T}^{H}$, are excellent observables to discriminate the Drell-Yan component from the gluon-fusion contribution using the $R^{\vh}$ double ratio. A particularly attractive feature of the double ratio is its sensitivity to new-physics effects that manifest themselves in distinct shape differences. An example is shown in \cref{fig:RVH_a}. This figure, made with \vhnnlo, reproduces a toy example from ref.~\cite{Harlander:2018yns} and compares the $R^{\vh}$ double ratio as a function of $M_{\vh}$ for the SM (blue line) with two modified versions of it. In the first scenario, the top-to-Higgs Yukawa coupling is forced to be zero, $y_t=0$ (green line), while in the second it is doubled with respect to the SM value, $y_t=2y_t^\mathrm{SM}$ (red line), where the SM value $y_t^\mathrm{SM}=m_t/v$ is defined as the ratio of the top-quark mass $m_t$ divided by the Higgs vacuum expectation value $v$. The pronounced impact of these modified couplings on the shape of the $R^{\vh}$ double ratio above the top-pair threshold, $M_{\zh} > 2m_t\approx 350\,\mathrm{GeV}$, originates from changes in the relative importance of the triangle and box diagrams which are depicted in \cref{fig:tri,fig:box}, respectively. It is well known that the interference between these two contributions with top-quark loops is significant in the SM and leads to a substantial reduction of the gluon-fusion cross section. When the top-to-Higgs Yukawa interaction is switched off, the destructive interference is absent and only the triangle diagram contributes, resulting in a sizable enhancement relative to the SM. Conversely, enhancing the top-to-Higgs Yukawa coupling strengthens the destructive interference near threshold, reducing the cross section in this region until the box contribution eventually dominates and causes the gluon-fusion prediction to exceed the SM at large invariant masses.

The \history{} framework allows for a straightforward extension of the $R^{\vh}$ double ratio to additional observables that have not yet been explored. As an example, \cref{fig:RVH_b} shows the $R^{\vh}$ double ratio as function of the Higgs rapidity, $y_H$, for the SM and the two modified Yukawa coupling scenarios. While the symmetry of the rapidity distribution is retained in both models, the shapes remain sensitive to the underlying physics. In the scenario without a direct top-to-Higgs Yukawa coupling, the $R^{\vh}$ double ratio decreases more rapidly with increasing absolute value of $y_H$ than in the SM, whereas the model with an enhanced top-to-Higgs Yukawa coupling exhibits a much flatter distribution. This makes $R^{\zh}(y_H)$ an additional promising observable in the search for possible BSM effects in the associated Higgs production process.

\section{Conclusions}

With the development of \history{}, we have created a powerful tool for studying fully-differential cross sections of inclusive color-singlet production processes at hadron colliders, both within the SM and in BSM scenarios, including QCD corrections up to NNLO accuracy. The fully-local NSC subtraction scheme is ideally suited for such computations, offering both numerical efficiency and the flexibility of a process-independent formalism. Although the \history{} framework has so far been applied only to the study of $\vh$ production through the quark-antiquark annihilation mechanism, extensions to other processes are planned. Moreover, we are actively working on the implementation of the subtraction terms required for color-singlet processes induced by gluons at Born level, which will remove the current limitations and further broaden the applicability of the framework.

Preliminary studies with \history{} for associated Higgs production in the SM, the Two-Higgs-Doublet Model, and the $B\!-\!L$ Model yield promising results. These findings motivate a more detailed investigation of this process and the $R^{\vh}$ double ratio in future work. Exploring additional observables, extending the analysis to further BSM scenarios, and achieving a deeper understanding of the correlations and uncertainties entering the $R^{\vh}$ double ratio could provide valuable guidance for experimental searches for new physics.

\acknowledgments

We thank Micha\l{} Czakon, Terry Generet, Robert Harlander, and Raoul Röntsch for numerous insightful discussions and for their constructive advice that contributed to the success of this project. We are also grateful to Robert Harlander for valuable comments on the manuscript.

This research was supported by the Deutsche Forschungsgemeinschaft (DFG, German Research Foundation) under grant 400140256, ``GRK 2497: The physics of the heaviest particles at the LHC'', and grant 396021762, ``TRR 257: P3H – Particle Physics Phenomenology after the Higgs Discovery''. S.Y.K.\@ receives funding from the Federal Ministry of Education and Research (BMBF) under grant number 05H21PACCA. The work of L.S.\@ is supported by the ERC (grant 101041109 ``BOSON''). Views and opinions expressed are however those of the authors only and do not necessarily reflect those of the European Union or the European Research Council Executive Agency. Neither the European Union nor the granting authority can be held responsible for them.

Computations were performed with computing resources granted by RWTH Aachen University under project rwth1826.

\bibliographystyle{shortref}
\bibliography{references}

\end{document}